\documentclass[aps,prl,twocolumn,twoside,superscriptaddress]{revtex4-1}
\usepackage{}

\usepackage{amsmath,amsthm,amssymb} %
\usepackage{color} %
\usepackage{csquotes} %
\usepackage{dsfont} %
\usepackage{times} 
\usepackage{graphicx} %
\usepackage{suffix} 
\usepackage{xspace} %
\usepackage{bbm} 
\usepackage{mathrsfs} 
\usepackage{lipsum}
\usepackage{mathtools}

\definecolor{dred}{rgb}{.8,0.2,.2}
\definecolor{ddred}{rgb}{.8,0.5,.5}
\definecolor{dblue}{rgb}{.2,0.2,.8}
\definecolor{dgreen}{rgb}{.2,0.5,.2}

\usepackage[colorlinks,linkcolor=dred,urlcolor=dblue,citecolor=dgreen]{hyperref}

\usepackage{tikz}
\usepackage{pgfplots}
\usetikzlibrary{shadows}

\newcommand*{\ket}[1]{\ensuremath{|#1\rangle}} %
\newcommand*{\bra}[1]{\ensuremath{\langle#1|}} %

\newcommand*{\norm}[1]{\ensuremath{\left\lVert #1 \right\rVert}} %
\newcommand*{\complex}{\mathbb{C}} %
\DeclareMathOperator{\tr}{tr} %
\DeclareMathOperator{\conv}{conv} %

\def\H{\mathcal{H}} %

\begin{document}


\title{A Separability-Entanglement Classifier via Machine Learning}

\author{Sirui Lu}
\thanks{These authors contributed equally to this work.}
\affiliation{Department of Physics, Tsinghua University, Beijing, 100084, China}

\author{Shilin Huang}
\thanks{These authors contributed equally to this work.}
\affiliation{Institute for Quantum Computing, University of Waterloo,
  Waterloo N2L 3G1, Ontario, Canada} %
\affiliation{Institute for Interdisciplinary Information Sciences, Tsinghua University, Beijing, 100084, China}

\author{Keren Li}
\affiliation{Department of Physics, Tsinghua University, Beijing, 100084, China}
\affiliation{Institute for Quantum Computing, University of Waterloo,
  Waterloo N2L 3G1, Ontario, Canada} %

\author{Jun Li}
\email{lijunwu@mail.ustc.edu.cn}
\affiliation{Institute for Quantum Computing, University of Waterloo,
  Waterloo N2L 3G1, Ontario, Canada} %
\affiliation{Beijing Computational Science Research Center, Beijing, 100193, China}
\affiliation{Department of Mathematics \& Statistics, University of
  Guelph, Guelph N1G 2W1, Ontario, Canada} %

\author{Jianxin Chen} %
\affiliation{Joint Center for Quantum Information and Computer
  Science, University of Maryland, College Park, Maryland, USA}

\author{Dawei Lu}
\email{d29lu@uwaterloo.ca}
\affiliation{Institute for Quantum Computing, University of Waterloo,
  Waterloo N2L 3G1, Ontario, Canada} %
  \affiliation{Department of Physics, Southern University of Science and Technology, Shenzhen 518055, China}
\affiliation{Department of Mathematics \& Statistics, University of
  Guelph, Guelph N1G 2W1, Ontario, Canada} %

\author{Zhengfeng Ji}%
\affiliation{Centre for Quantum Computation \& Intelligent Systems,
  School of Software, Faculty of Engineering and Information
  Technology, University of Technology Sydney, Sydney, Australia}%
\affiliation{State Key Laboratory of Computer Science, Institute of
  Software, Chinese Academy of Sciences, Beijing, China}%

\author{Yi Shen}%
\affiliation{Department of Statistics and Actuarial Science, University of Waterloo,
  Waterloo N2L 3G1, Ontario, Canada}%

\author{Duanlu Zhou}%
\affiliation{Institute of Physics, Chinese Academy of Sciences,
  Beijing 100190, China}%

\author{Bei Zeng} %
\affiliation{Department of Mathematics \& Statistics, University of
  Guelph, Guelph N1G 2W1, Ontario, Canada} %
\affiliation{Institute for Quantum Computing, University of Waterloo,
  Waterloo N2L 3G1, Ontario, Canada} %
    \affiliation{Department of Physics, Southern University of Science and Technology, Shenzhen 518055, China}

\begin{abstract}
The problem of determining whether a given quantum state is entangled lies at the heart of quantum information processing, which is known to be an NP-hard problem in general. Despite the proposed many methods – such as the positive partial transpose (PPT) criterion and the $k$-symmetric extendibility criterion – to tackle this problem in practice, none of them enables a general, effective solution to the problem even for small dimensions. Explicitly, separable states form a high-dimensional convex set, which exhibits a vastly complicated structure. In this work, we build a new separability-entanglement classifier underpinned by machine learning techniques. Our method outperforms the existing methods in generic cases in terms of both speed and accuracy, opening up the avenues to explore quantum entanglement via the machine learning approach.
\end{abstract}

\date{\today}


\maketitle

Born from pattern recognition, machine learning possesses the capability to make decisions without being explicitly programmed after learning from large amount of data.  Beyond its extensive applications in industry, machine learning has also been employed to investigate physics-related problems in recent years.  A number of promising applications have been proposed to date, such as the Hamiltonian learning~\cite{WGFC14},  automated quantum experiments generation~\cite{KMFLZ16}, identification of phases and phase transition~\cite{SCSKL16,NLH17,CM17}, efficient representation of quantum many-body states \cite{taoxiang17,Duan17},  just to name a few. Nevertheless, there are yet a myriad of significant but hard problems in physics to be assessed, in which should machine learning provide more novel insights.
For example, to determine whether a generic quantum state is entangled or not is a fundamental and NP-hard problem in quantum information processing \cite{gurvits2003classical}, and machine learning is demonstrated to be exceptionally effective in tackling it as shown in this work.

As one of the key features in quantum mechanics, entanglement allows
two or more parties to be correlated in a way that is much stronger than they
can be in any classical way~\cite{Horodecki2009}.  It also plays a key role  in many quantum information processing tasks such as teleportation and quantum key distribution~\cite{NC00}. As a result, one question naturally arises: is there a universal criterion to tell if an arbitrary  quantum state is separable or entangled?  This is a typical classification problem, which remains of great challenge even for bipartite states. In fact, such an entanglement detection problem is proved to be NP-hard \cite{gurvits2003classical}, implying that it is almost impossible to devise an efficient algorithm in complete generality.

Here, we focus on the task of detecting bipartite entanglement.
Consider a bipartite system $AB$ with the Hilbert space  $\mathcal{H}_A \otimes \mathcal{H}_B$,
where $\mathcal{H}_A$ has dimension $d_A$ and $\mathcal{H}_B$ has dimension $d_B$, respectively.
A state $\rho_{AB}$ is separable if it can be written as a convex combination $\rho_{AB} = \sum_i \lambda_i \rho_{A,i}\otimes \rho_{B,i}$ with a probability distribution $\lambda_i\geq 0$ and $\sum_i \lambda_i=1$. Here $\rho_{A,i}$ and
$\rho_{B,i}$ are density operators acted on $\H_A$, $\H_B$ respectively. Otherwise, $\rho_{AB}$ is entangled~\cite{werner1989quantum,guhne2009entanglement}. To date, many criteria have been proposed to detect bipartite entanglement, each with its own pros and cons. For instance, the most famous criterion is the positive partial transpose (PPT) criterion, saying that a separable state must have PPT;
however, it is only necessary and sufficient when $d_Ad_B \le 6$
\cite{Per96,HHH96}.   Another widely used one is the $k$-symmetric extension hierarchy \cite{navascues2009power,brandao2011quasipolynomial},  which is presently one of the most powerful criteria, but   hard to compute in practice due to its exponentially growing complexity with $k$ \cite{S}.

\begin{figure}[t]
\begin{center}
\includegraphics[width=0.95\linewidth]{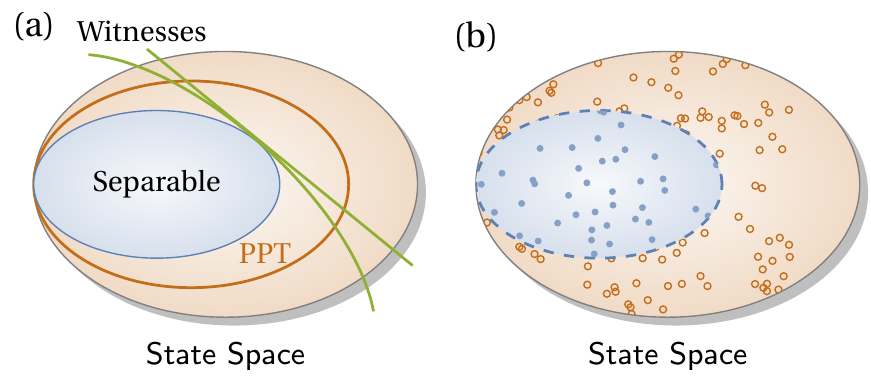}
\end{center}
\caption{(a) In the high-dimensional space, the set of all states
is convex, while separable states form a convex subset.  Many criteria, such as the linear (green straight line) or nonlinear (green curve) entanglement witnesses \cite{GL06,Horodecki2009} and  PPT tests, are based on this geometric structure and detect a limited set of entangled states. (b) Classifier built from supervised   learning has a decision boundary of highly complex shape.}
\label{classifier}
\end{figure}

In this work, we employ the machine learning techniques to tackle the bipartite entanglement detection problem by recasting it as a learning task, namely we attempt to construct a separability-entanglement classifier. Due to its renowned effectiveness in pattern recognition for high-dimensional objects, machine learning is a powerful tool to solve the above problem. In particular, a reliable separability-entanglement classifier in terms of speed and accuracy is constructed via the supervised learning approach.  The idea is to feed our classifier by a large amount of  sampled trial states as well as their corresponding class labels (separable or entangled), and then train the classifier to predict the class labels of new states that it has not encountered before. It is worthy stressing that,
there is also a remarkable improvement with respect to universality in our classifier compared to the conventional methods. Previous methods only detect a limited part of the state space,  e.g. different entangled states often require different entanglement witnesses. In contrast, our classifier can handle a variety of  input states once properly trained, as shown in Fig. \ref{classifier}.

\emph{Supervised learning} -- The bipartite entanglement detection problem can be formulated as a supervised binary classification task.
Following the standard procedure of supervised learning \cite{MRT10,Shai14},
the feature vector representation of the input objects (states) in a bipartite system \emph{AB} is first created.
Indeed, any quantum state $\rho$, as a density operator acting on  $\mathcal{H}_A \otimes \mathcal{H}_B$
can be represented as a real vector $\bm x$ in $\mathscr{X} = \mathbb{R}^{d_{A}^{2} d_{B}^{2}-1}$, which is due to the fact that $\rho$ is Hermitian and of trace $1$ (see Supplementary Material \cite{S}).
In the machine learning language,
we refer $\bm{x}$ as the feature vector of $\rho$ and $\mathscr{X}$   the feature space.

Next, a dataset of training examples is produced, with the form $\mathcal{D}_{\text{train}} = \left\{ (\bm{x}_1, y_1), ..., (\bm{x}_n, y_n) \right\}$, 
where $n$ is the size of the set, $\bm{x}_i \in \mathscr{X}$ is the $i$-th sample, and $y_i$ is its corresponding label signifying which class it belongs to, i.e., $y_i$ equals to 1 if $\bm{x}_i$ is entangled or $-1$   otherwise. 
When $d_Ad_B \le 6$, the labeling process can be  directly computed via the PPT criterion.
For higher-dimensional cases, we attempt to estimate the labels by   convex hull approximation, which we will   describe  later.
The task is to analyze these training data and produce an inferred classifier   that predicts the unknown class labels for generic new input states. 

Explicitly, the aim of supervised learning is to infer a function (classifier) $h: \mathscr{X} \to \left\{-1, 1\right\}$ 
among a fixed class of functions $\mathscr{H}$ such that $h$ is expected to be close to the true decision function.
One basic approach to choose $h$ is the so-called empirical risk minimization, 
which seeks the function that best fits the training data among the class $\mathscr{H}$. 
In particular, to evaluate how well $h$ fits the training data $\mathcal{D}_{\text{train}}$, a loss function is defined as
\begin{equation}
\mathcal{L}(h, \mathcal{D}_{\text{train}}) = \frac{1}{|\mathcal{D}_{\text{train}}|}\sum_{(\bm{x}_i, y_i)\in \mathcal{D}_\text{train}} \mathbbm{1}(y_i \neq h(\bm{x}_i)),
\label{loss}
\end{equation}
where $\mathbbm{1}(\cdot)$ is the truth function of its arguments. 
For a  generic new input test dataset $\mathcal{D}_\text{test}$ that  contains previously unseen data, function $\mathcal{L}(h, \mathcal{D}_\text{test})$ gives a quantification of the generalization error from $\mathcal{D}_\text{train}$ to $\mathcal{D}_\text{test}$.

Numerous supervised learning algorithms have been developed, each with its strength and weakness. These algorithms, which have distinct choices of class $\mathscr{H}$, include support vector machine (SVM) \cite{cortes1995support},  decision trees \cite{breiman1984classification}, bootstrap aggregating \cite{Zhou12},  and boosting \cite{schapire2003boosting}, etc.
We have applied these algorithms to the separability problem directly,
but neither of them provided an acceptable accuracy, which is mainly due to the lack of prior knowledges for training, e.g., the geometric shape of the set of separable states $\mathcal{S}$. 
Taking the kernel SVM approach~\cite{cortes1995support} as an example, it uses a kernel function to map data from the original feature space to another Hilbert space, and then finds a hyperplane in the new space to split the data into two subclasses. 
It turns out that using common kernels such as radial basis function and polynomials, the error rate on the test dataset is always around $10\%$ (see Supplementary Material~\cite{S} for details).
This suggests that the boundary of  $\mathcal{S}$ is too complicated to be portrayed by manifolds with ordinary shapes.

\emph{Convex hull approximation} -- 
The above discussions suggest that it is desirable to examine the detailed geometric shape of $\mathcal{S}$ in advance.
One well-known approach is to approximate $\mathcal{S}$ from outsize via $k$-symmetric extendible set $\Theta_k$, where $\Theta_k \supset \Theta_{k+1}$ and $\Theta_k$ converges exactly to $\mathcal{S}$ as $k$ goes to infinity~\cite{doherty02a}.
Unfortunately, it is impractical to compute the boundary of $\Theta_k$ for large $k$, while it is still far from approximating $\mathcal{S}$ for small $k$~\cite{S}. 

\begin{figure}[b]
\begin{center}
\includegraphics[width=0.95\linewidth]{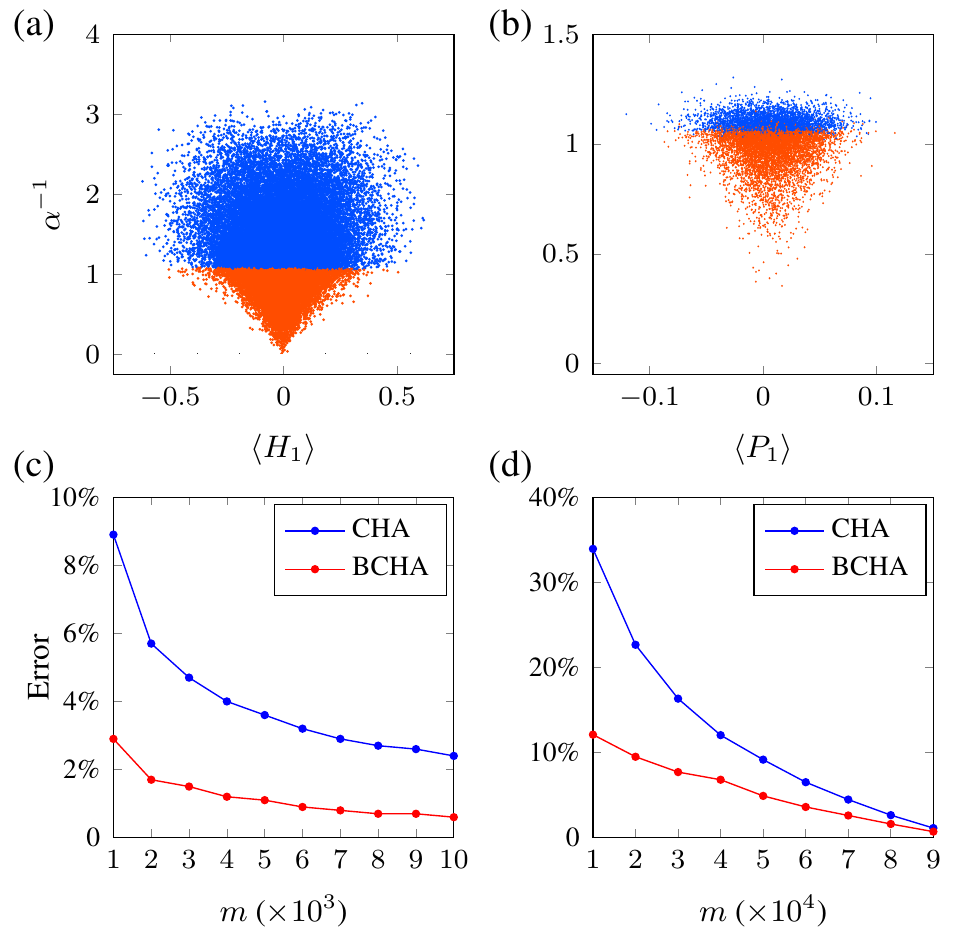}
\end{center}
\caption{Results of the CHA approach and BCHA classifier. More details of the BCHA classifier can be found in \cite{S}.
(a)
Test results of BCHA when $m = 10^3$ for the two-qubit case.
The random density matrices in the test dataset are projected on a plane by projection
$\pi: (\bm{x},\alpha) \rightarrow (x_1, \alpha^{-1})$. Here $H_1 =  | 0\rangle\langle 0| \otimes \sigma_z /\sqrt{2} $. 
The red points are states predicted as separable, while the blue points are states predicted as entangled.
Some states with $\alpha < 1$ are predicted as separable, which is different from CHA.
(b) Test results of BCHA when $m = 2\times 10^4$ for the two-qutrit case. Here $P_1 = (|00\rangle \langle 00| - |01\rangle \langle 01|)\sqrt{3}/2$.
All the states in the test dataset are PPT states. The red points are states predicted as separable, while the blue points are states predicted as bound entangled. 
(c) Comparison between CHA and BCHA for two-qubit states. For the same $m$,   BCHA   clearly suppresses the
error rate significantly. And to achieve the same error rate,   BCHA requires much less running time (which   mainly depends on the value of $m$). For instance, to decrease the error rate to less than $3\%$, CHA requires a convex hull with $m \approx 7 \times 10^3$, while BCHA only requires a convex hull with $m \approx 10^3$, which considerably reduces the computational cost.
(d) Comparison between CHA and   BCHA for two-qutrit PPT states.  From the data, similar to the two-qubit case, BCHA also outperforms CHA in terms of speed and accuracy. 
}
\label{result}
\end{figure}

However, it is much easier to approximate $\mathcal{S}$ from inside, 
since $\mathcal{S}$ is a closed convex set, 
and   its extreme points are exactly all   the separable pure states, which can be straightforwardly parameterized and generated numerically.
We randomly sample $m$   separable pure states $\bm{c}_1,...,\bm{c}_m \in \mathscr{X}$ to form a convex hull $\mathcal{C} \coloneqq \conv(\left\{ \bm{c}_1,...,\bm{c}_m \right\})$. $\mathcal{C}$ is said to  be a convex hull approximation (CHA) of $\mathcal{S}$,
with which we can approximately tell whether a state $\rho$ is separable or not by testing if its feature vector $\bm{p}$ is in $\mathcal{C}$. 
This is equivalent to determining whether $\bm{p}$ can be written as a convex combination of $\bm{c}_i$ by solving the following linear programming:

\begin{figure}[t]
\begin{center}
\includegraphics[width=0.975\linewidth]{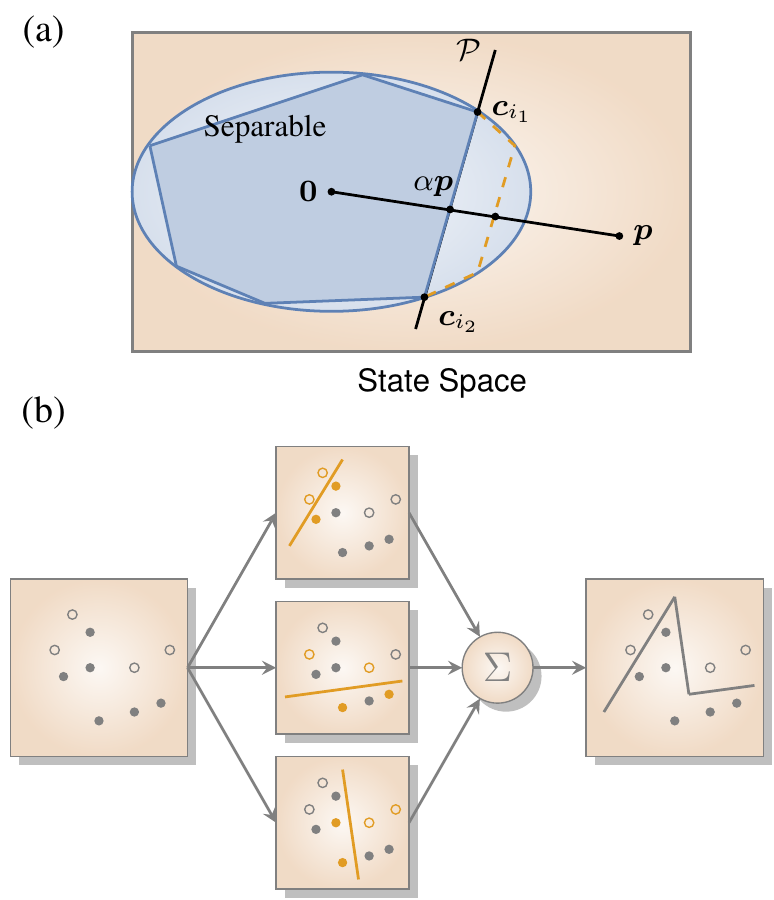}
\end{center}
\caption{
(a) Illustration of the iterative algorithm for detecting the separability.
Initially we build a CHA $\mathcal{C}$.
For a state $\rho$ with feature vector $\bm{p}$, we find the maximum $\alpha$ such that $\alpha\bm{p}$ is still in $\mathcal{C}$.
If $\alpha \ge 1$, then $\rho$ is surely separable. 
Otherwise, suppose $\alpha \bm{p}$ lies on a hyperplane $\mathcal{P}$, such that $\mathcal{P} \cap \mathcal{C}$ is the boundary of $\mathcal{C}$.
Let $\bm{c}_{i_1}, \ldots, \bm{c}_{i_D}$ be extreme points of $\mathcal{C}$ that are in $\mathcal{P} \cap \mathcal{C}$.
We enlarge $\mathcal{C}$ by sampling separable pure states that are near $\bm{c}_{i_1}, \ldots, \bm{c}_{i_D}$, then repeat the above procedure for many times, until $\alpha \ge 1$ or $\alpha$ converges.
(b) Illustration of the learning algorithm: ensemble methods are models composed of multiple weaker models that are independently trained and whose predictions are combined in some way to make the overall prediction.
For example, each time we draw a subset of training data (marked as yellow dots in the figure), 
and we train a model based on this subset, which is a weak model on the whole training set. 
We repeat the process for many times and obtain a batch of weak models, and combine them as a committee.
}
\label{method}
\end{figure}

\begin{align}
\max \quad  & \alpha \quad \text{s.t} \quad \alpha\bm{p} \in \mathcal{C},\nonumber \\
\text{i.e.} \quad & \alpha \bm{p} =\sum_{i=1}^m\lambda_i \bm{c}_i , \quad \lambda_i \ge 0,\quad \sum\nolimits_{i} \lambda_i = 1.
\label{optimize}
\end{align}
Here $\alpha = \alpha(\mathcal{C}, \bm{p})$ is a function of $\mathcal{C}$ and $\bm{p}$. 
If $\alpha(\mathcal{C}, \bm{p}) \ge 1$, $\bm{p}$ is in $\mathcal{C}$ and thus $\rho$ is separable; otherwise, $\rho$ is highly possible to be an entangled state.
In principle, $\mathcal{C}$ will be a more accurate CHA of $\mathcal{S}$ if we construct $\mathcal{C}$ with more extreme points.
We test the error rate of CHA on a set of $2 \times 10^4$ random two-qubit states, which is sampled under a specified distribution ~\cite{S} and labeled by PPT criterion. 
The results are shown by the blue curve in Fig.~\ref{result}(c), where the error rate decreases quickly to $3\%$ when the number of extreme points $m$ increases to $10^4$.

However, we can not directly test the accuracy of CHA on generic two-qutrit states, since PPT criterion is no longer sufficient for detecting separability.
To illustrate the power of CHA beyond the PPT criterion, we use a specific example that is previously well-studied.
Consider a set of two-qutrits pure states $\{\ket{v_1},\dots,\ket{v_5}\}$ that form the well-known unextendible product basis~\cite{bengtsson2007geometry}, where
$\ket{v_1}=(\ket{00} -\ket{01} )/\sqrt{2}$,
$\ket{v_2}=(\ket{21} -\ket{22}) /\sqrt{2}$,
$\ket{v_3}=(\ket{02} -\ket{12}) /\sqrt{2}$,
$\ket{v_4}=(\ket{10} -\ket{20}) /\sqrt{2}$, and
$\ket{v_5}=(\ket{0}+\ket{1}+\ket{2})^{\otimes 2}/3$.
It is known that 
$\rho_{\text{tiles}} = (\mathbb{I}-\sum_{i=1}^{5}\ket{v_i}\bra{v_i} )/4$ is an entangled state with PPT~\cite{bennett1999unextendible}. 
Due to the fact that $\mathcal{S}$ is convex and closed, there must exist a unique critical point $\alpha_{\text{tiles}} \in [0,1)$ such that 
$\alpha_{\text{tiles}} \rho + (1 - \alpha_{\text{tiles}}) \mathbb{I} / (d_Ad_B)$,
the probabilistic mixture of $\rho$ and the maximally-mixed state $\mathbb{I}/(d_Ad_B)$, is on the boundary of $\mathcal{S}$.
Ref. \cite{johnston2014detection} compared the effectiveness of various separability criteria, 
and   concluded that $\alpha_{\text{tiles}} \in (0.5643, 0.8649]$. 
Note that for a CHA $\mathcal{C}$, $\alpha(\mathcal{C}, \bm{p}_{\text{tiles}})$ actually provides a lower bound approximation of $\alpha_{\text{tiles}}$,
where $\bm{p}_{\text{tiles}}$ is the feature vector of $\rho_{\text{tiles}}$.
Now we   apply CHA and attempt to improve the lower bound of $\alpha_{\text{tiles}}$, and the result is shown in Table~\ref{tiles}.

\begin{table}[t]
\centering
{\footnotesize\renewcommand{\arraystretch}{1.5}\setlength{\tabcolsep}{5pt}
\begin{tabular}{ccccccc}
\hline
$m$   & 2000 & 5000 & 10000 & 20000 & 50000 & 100000 \\
$\alpha(\mathcal{C},\bm{p}_{\text{tiles}})$ &  0.5264 & 0.5868 & 0.6387  & 0.6759 & 0.7150 & 0.7459 \\ \hline
\end{tabular}
}
\caption{Numerical results for approxmiating $\alpha_\text{tiles}$ by $\alpha(\mathcal{C},\bm{p}_\text{tiles})$. Here, $m$ is the number of random extreme points for building $\mathcal{C}$. }
\label{tiles}
\end{table}

We find that the lower bound of $\alpha_{\text{tiles}}$ has been raised to $0.7459$.  However, in Table~\ref{tiles}, the value of $\alpha(\mathcal{C}, \bm{p}_{\text{tiles}})$ has not converged yet. To reach the convergence of $\alpha(\mathcal{C}, \bm{p}_{\text{tiles}})$, we have to enlarge $\mathcal{C}$ by adding more extreme points.
However, note that the point $\alpha(\mathcal{C}, \bm{p}_{\text{tiles}})$ lies on a part of the boundary of $\mathcal{C}$, which is the intersection of a hyperplane and $\mathcal{C}$. 
Let $\bm{c}_{i_1}, \ldots, \bm{c}_{i_D}$ be the extreme points of $\mathcal{C}$ that lie on the hyperplane as well.
Clearly, if we enlarge $\mathcal{C}$ by sampling the separable pure states that are near $\bm{c}_{i_1}, \ldots, \bm{c}_{i_D}$ rather than
sampling uniformly over the whole set of separable pure states,
it will boost the value of $\alpha(\mathcal{C}, \bm{p}_{\text{tiles}})$ more effectively.

Subsequently, we refine CHA as an iterative algorithm ~\cite{S}, with the idea shown in Fig. \ref{method}(a). The iterative algorithm gives the result $\alpha_{\text{tiles}} > 0.8648$. 
As the upper bound of $\alpha_\text{tiles}$ is $0.8649$ ~\cite{johnston2014detection}, we can explicitly conclude that $\alpha_{\text{tiles}} \approx 0.8649$.
It is worthy emphasizing that the algorithm also gives the critical point for a generic entangled state with small error, and detects the separability for generic separable states~\cite{S}. 

\emph{Combining CHA and supervised learning} -- There is yet a noticeable drawback of the above CHA approach from the perspective of the tradeoff between the accuracy and time consumption. Boosting the accuracy means adding additional extreme points to enlarge the convex hull, which leads to more time costs to
determine if a point is inside the enlarged convex hull or not. To overcome this, we combined CHA with supervised learning, as machine learning has the power to speed up such computations.

To design a learning process that is suitable for our problem,
for each state $\rho$ with feature vector $\bm{p}$, we extend the feature vector as $(\bm{p}, \alpha(\mathcal{C}, \bm{p}))$ 
in order to encode the geometric information of the CHA $\mathcal{C}$ into the dataset.
In this manner, the training dataset is written as $\mathcal{D}_{\text{train}} = \left\{ (\bm{x}_1, \alpha_1, y_1), ..., (\bm{x}_n, \alpha_n, y_n) \right\}$, where $\alpha_i = \alpha(\mathcal{C}, \bm{x}_i)$. A classifier $h$ is now a binary function defined on $\mathscr{X} \times \mathbb{R}$, and the loss function of a classifier $h$ is then redefined as
\begin{equation}
\mathcal{L}(h, \mathcal{D}_{\text{train}}) = \frac{1}{|\mathcal{D}_{\text{train}}|}
\sum_{\left(\bm{x}_i, \alpha_i, y_i\right)\in \mathcal{D}_{\text{train}}} \mathbbm{1}(y_i \neq h(\bm{x}_i, \alpha_i)).
\end{equation}
Subsequently, we employ a standard ensemble learning approach \cite{Zhou12} to train a classifier with training data $\mathcal{D}_{\text{train}}$, described as follows.

The essential idea of ensemble learning is to improve the predictive performance by combining multiple classifiers into a committee. Even though the prediction from each constituent might be poor,   the combined classfier   could often still perform excellent.
For the binary classification problem, we can train different classifiers to give their respective binary votes for each prediction, and use the majority rule
to choose the value which receives more than half votes as the final answer, see Fig. \ref{method}(c) for a schematic diagram.

Here, we choose bootstrap aggregating (bagging) \cite{breiman1996bagging} as our training ensemble algorithm.
In each run, a training subset is randomly drawn from the whole set $\mathcal{D}_{\text{train}}$, 
and a model is trained from the training subset using another learning algorithm, e.g., decision trees learning. 
We repeat the process for $L = 100$ times and obtain $L$ different models, 
which are finally combined together as the committee. 
Since $\alpha_i = \alpha(\mathcal{C}, \bm{x}_i)$ contains the geometric information of CHA $\mathcal{C}$, our method is indeed a combination of bagging and CHA. We call this combined method BCHA.

The computational cost contains two parts: the cost of computing $\alpha_i$ via linear programming,
and the time of computing each constituent in the committee. The latter cost is much smaller than the former. 
Therefore, by using a convex hull of much smaller size and implementing a bagging algorithm, 
a significant boost in terms of accuracy is anticipated if the total computational cost is fixed. 
For the two-qubit case, we have demonstrated such a remarkable boost of accuracy in our BCHA classifier, as shown in Fig. \ref{result}(c), where the advantages of the BCHA classifier in terms of both accuracy and speed are shown.

We further extend the classifier to the two-qutrit scenario. Unlike the two-qubit case, the critical question now is how to set an appropriate criterion to evaluate whether the classifier is working correctly, since PPT criterion is not sufficient for detecting separability in two-qutrit systems. As the convex hull is capable of approximating the set of separable
states $\mathcal{S}$ to an arbitrary precision, we use $10^5$ random separable pure states as extreme points to form the hull, and assumed
it to be the true $\mathcal{S}$.
The learning procedure is analogous to the one used for two qubits.
Figure \ref{result}(d) shows the accuracy of the BCHA classifier compared to that of the sole CHA approach, 
Similar as the two-qubit case, the BCHA classifier shows clear advantage in terms of both accuracy and speed in comparison with the sole CHA method.

\textit{Conclusion} -- In summary, we study the entanglement detection problem via the machine learning approach, and build a reliable separability-entanglement classifier by combining supervised learning and the CHA method. Compared to the conventional criteria for entanglement detection, our method can classify an unknown state into the separable or entangled category more precisely and rapidly. The classifier can be extended to higher dimensions in principle, and the developed techniques in this work would also be incorporated in future entanglement-engineering experiments. We anticipate that our work would provide new insights to employ the machine learning techniques to deal with more quantum information processing tasks in the near future.

\begin{acknowledgements}
\textit{Acknowledgments} -- 
This work is supported by Chinese Ministry of Education under grants No.20173080024.
We thank Daniel Gottesman and Nathaniel Johnston for helpful discussions. JL is supported by the National Basic Research Program of China (Grants No. 2014CB921403, No. 2016YFA0301201, No. 2014CB848700 and No. 2013CB921800), National Natural Science Foundation of China (Grants No. 11421063, No. 11534002, No. 11375167 and No. 11605005), the National Science Fund for Distinguished Young Scholars (Grant No. 11425523), and NSAF (Grant No. U1530401). DL and BZ are supported by NSERC and CIFAR.
JC was supported by the Department of Defense (DoD).
DZ is supported by NSF of China (Grant No. 11475254), BNKBRSF
of China (Grant No. 2014CB921202), and The National Key Research and
Development Program of China(Grant No. 2016YFA0300603).
\end{acknowledgements}


%

\appendix

\section{A. Generalized Gell-Mann Matrices}

To represent a $n$-by-$n$ density matrix $\rho$ as a real vector $\bm{x}$ in $\mathbb{R}^{n^2-1}$, we can find a Hermitian orthogonal basis that contains identity such that $\rho$ can be expanded in such a basis with real coefficients. For example, the Pauli basis is a commonly used one.
In our numerical tests, we take the generalized Gell-Mann matrices and the identity as the Hermitian orthogonal basis.
In this section, we recall the definition of the generalized Gell-Mann matrices, which is shown in~\cite{bertlmann2008bloch}.

Let $\left\{ \ket{1}, \ldots, \ket{n} \right\}$ be the computatioal basis of the $n$-dimensional Hilbert space, and $E_{j,k} = \ket{j}\bra{k}$.
We now define three collections of matrices. The first collection is symmetric:
\[
	s_{j,k} = E_{j,k} + E_{k,j}
\]
for $1 \le j < k \le n$. The second collection is antisymmetric:
\[
	a_{j,k} = -i\left(E_{j,k} - E_{k,j}\right)
\]
for $1 \le j < k \le n$. The last collection is diagonal:
\[
	d_{l} = \sqrt{\frac{2}{l(l+1)}} \left( \sum_{j=1}^{l} E_{j,j} - l E_{l+1,l+1} \right)
\]
for $1 \le l \le n-1$.

The generalized Gell-Mann matrices are elements in the set $\left\{\lambda_i\right\} = \left\{s_{j,k}\right\} \cup \left\{a_{j,k}\right\} \cup \left\{ d_l \right\}$, which gives a total of $n^2-1$ matrices.
We can easily check that $$\text{tr}\left(\lambda_i\right) = \text{tr}\left(\lambda_i \mathbb{I} \right) = 0$$ and $$\text{tr}\left(\lambda_i \lambda_j\right) = 2 \delta_{ij},$$
which implies that $\left\{\lambda_i\right\} \cup \left\{\mathbb{I}\right\}$ forms an orthogonal basis of observables in $n$-dimensional Hilbert space.

For every $n$-by-$n$ density matrix $\rho$, $\rho$ can be expressed as a linear combination of $\lambda_i$ and $\mathbb{I}$ as follows:
\[
  \rho=\frac{1}{n}\left(\mathbb{I}+\sqrt{\frac{n(n-1)}{2}}\bm{x}\cdot\vec{\lambda}\right),
\]
where $\bm{x}=(x_1,x_2,\dots,x_{n^2-1})\in \mathbb{R}^{n^2-1}$ satisfies
$$x_i = \sqrt{\frac{n}{2(n-1)}}\text{tr}\left(\rho \lambda_i\right).$$

\section{B. The set of $k$-extendible states}

In this section, we recall facts regarding $k$-extendible states and its relationship to separability.

A bipartite state $\rho_{AB}$ is said to be $k$-symmetric extendible
if there exists a global state $\rho_{AB_1\ldots B_k}$ whose reduced density matrices $\rho_{AB_i}$ are equal to $\rho_{AB}$ for $i = 1, \ldots, k$.
The set of all $k$-extendible states, denoted by $\Theta_k$, is convex with a hierarchy structure $\Theta_k \supset \Theta_{k+1}$.
Moreover, when $k \rightarrow \infty$, $\Theta_k$ converges exactly to the set of separable states~\cite{doherty02a}.

The $\Theta_k$ is known
to be closely related to the ground state of some $(k+1)$-body Hamiltonians~\cite{zeng2015quantum}.
To be more precise, consider a $2$-local Hamiltonian $H$ of a $(k+1)$-body system with Hilbert space
$\mathbb{C}^{d_A}\bigotimes_{i=1}^k\mathbb{C}^{d_{B_i}}$ of dimension $d_Ad_B^{k}$, as given in the following form
$
H=\sum_{i=1}^k H_{AB_i}.
$
Here $H_{AB_i}$ is any Hermitian operator acting nontrivially on particles $A$ and $B_i$, and trivially on other $k-1$ parties. In other words, we will have $H_{AB_1}=H_{AB}\otimes I_{2,\ldots,k}$ ($I_{2,\ldots,k}$ is the identity operator of $B_2,\ldots,B_k$), and given the symmetry of $B_i$s, we can always write the nontrivial action of $H_{AB_i}$ on $\mathbb{C}^{d_A}\bigotimes_{i=1}^k\mathbb{C}^{d_{B_i}}$ in terms of some $H_{AB}$ acted on $d_Ad_B$-dimensional Hilbert space.

For any given $H$, denote its normalized ground state by $\ket{\psi_g}\in\mathbb{C}^{d_A}\bigotimes_{i=1}^k\mathbb{C}^{d_{B_i}}$, and $\rho_g=\ket{\psi_g}\bra{\psi_g}$.
Then the extreme points of $\Theta_k$ are given by the marginals of $\rho_{g}$ on particles $AB_i$, which are the same for any $i$. Denote this marginal by $\rho_{H}$, since it is completely determined by $H$.

To generate random extreme points of $\Theta_k$, we will need to first parametrize them.  Denote $\{O_{AB}^{lm}\}$ as set of orthonormal Hermitian basis for operators on $H_A\otimes H_{B_j}$ (see section A), then we can always write
$
H_{AB}=\sum_{lm} a_{lm} O_{AB}^{lm},
$
with parameters $a_{lm}$.
Without loss of generality, we assume $O_{AB}^{00}=I$,
and we will assume $a_{00}=0$, so there are only $d_A^2d_B^2-1$
terms in the sum. Since $H_{AB}$ is a Hermitian
matrix, $a_{lm}$ can be chosen as real, and we can further require that
$
\sum_{l,m} a_{lm}^2=1.
$
Consequently, $\rho_H$ will be a point in $\mathbb{R}^{d_A^2 d_B^2-1}$,
which is parametrized by $\{a_{lm}\}$. And the coordinate of $\rho_H$
are explicitly given by $b_{lm}=\tr(\rho_H O_{AB}^{lm})$.

Also, each $H_{AB}$ gives an entanglement witness.
The ground state energy of $H_{AB}$ is given by
$
E_0=\bra{\psi_g}H\ket{\psi_g}=\sum_{lm} a_{lm}b_{lm}/k.
$
For any density matrix $\rho_{AB}$,
if $\tr(\rho_{AB}H_{AB})<E_0$, then $\rho_{AB}$ has no $k$-symmetric extension,
hence is surely entangled.

\begin{figure}[t]
\begin{center}
\includegraphics[width=0.95\linewidth]{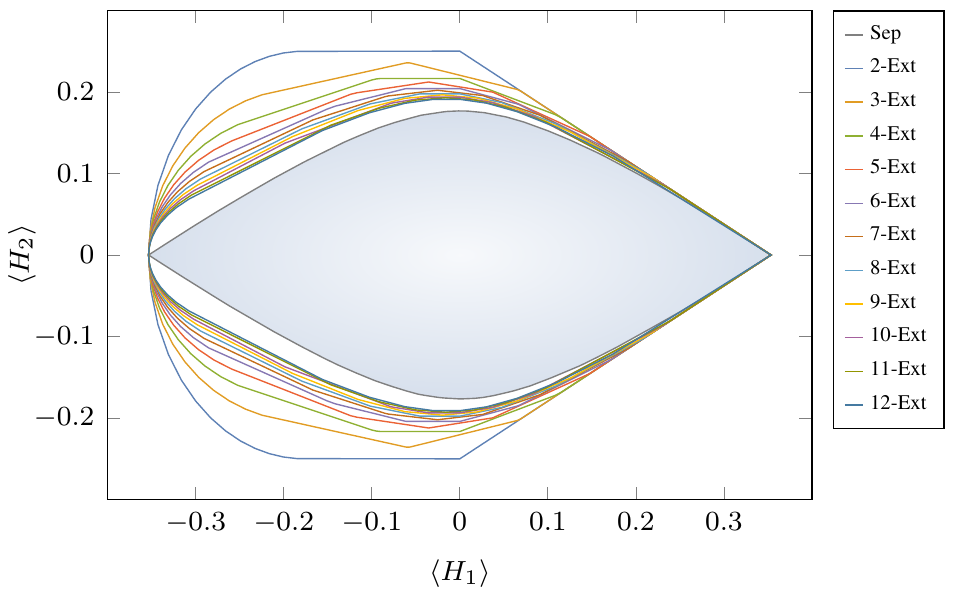}
\end{center}
\makeatletter
\renewcommand{\thefigure}{S\@arabic\c@figure}
\makeatother
\caption{Projections of the boundaries of the separable states and the
$k$-symmetric extendible states ($k=2,...,12$) on the plane spanned by the operators
$H_1 =  | 0\rangle\langle 0| \otimes \sigma_z /\sqrt{2} $ and $H_2 = (\sigma_y\otimes\sigma_x - \sigma_x\otimes\sigma_y)/2$.
Here $\sigma_x,\sigma_y,\sigma_z$ are the three Pauli operators. As we can see, there is still a large gap between $\Theta_{12}$ and separable set.}
\label{kextension}
\end{figure}

Since the dimension of $H$ grows exponentially with
$k$, to generate these extreme points for $\Theta_k$ becomes hard when $k$ increases.
In practice, we can generate the extreme points of $\Theta_{k}$ for $k = 12$ and $d_A = d_B = 2$.
However, as depicted in Fig. \ref{kextension}, there is still a large gap between the separable boundary and the $k$-extension boundary $k$.

More general properties on $k$-extendability and
its relationship to the quantum marginal problem can be found in Refs. ~\cite{Kly06,Col63,Erd72,Liu06,LCV07,WMN10}.

\section{C. Generating Random Density Matrices}

Since our aim is to determine the separability of generic bipartite states,
we require a bunch of random density matrices with full rank to test the performance of our approaches.
In our numerical tests, we sample random density matrices under the probability measure
$\mu = \nu \times \Delta_{\lambda}$,
where $\nu$ is the uniform distribution on $U(n)$ according to Haar measure, $\Delta_{\lambda}$ is the Dirichlet distribution
on the simplex $\sum_{i=1}^{n} d_i = 1$. The probability density function of Dirichlet distribution is
$$\Delta_{\lambda}\left(d_1, \ldots, d_n\right) = C_{\lambda}\prod_{i=1}^{n} d_i^{-\lambda},$$
where $\lambda > 0$ is a parameter and $C_{\lambda}$ is the normalization constant. Since every density matrix is unitarily similar to
a real diagonal density matrix, $\mu$ is a probability measure on the set of all density matrices. Such a probability measure is
discussed in Ref. \cite{zyczkowski1999volume}, section II.A.
We implemented the sampling on $\nu$ via directly calling the function \href{http://www.qetlab.com/RandomUnitary}{\textbf{\texttt{RandomUnitary}}}
in~\cite{johnston2016qetlab}. The entire implementation is in the code \href{http://qmlab.org/downloads/}{\textbf{\texttt{RandomState.m}}} on the website of QMLab~\cite{qmlab}.

For the 2-qubit case, we set $\lambda = 1/2$ and generate $5 \times 10^4$ random quantum states, which are put in the file \href{}{\textbf{\texttt{2x2rdm.mat}}}.
We have found that $35\%$ of the states are PPT states, i.e., separable states, which is consistent with the result shown in \cite{zyczkowski1999volume}.

For the 2-qutrit case, we also set $\lambda = 1/2$. As shown in~\cite{zyczkowski1999volume}, only $2.2\%$ of the random states are PPT states when $\lambda = 1/2$. However, our main interest is determining whether a PPT state is entangled. Thus, we reject all the states with negative partial transpose while sampling,
and obtain $2 \times 10^4$ PPT states eventually. These states are put in the file \href{http://qmlab.org/downloads}{\textbf{\texttt{3x3rdm.mat}}} on~\cite{qmlab}.
We can verify that at least $66.24\%$ of the PPT states are separable under the probability measure we have chosen, using the convex hull approximation, which will be discussed in section E.

We determine whether a state is a PPT state via \href{http://www.qetlab.com/IsPPT}{\textbf{\texttt{IsPPT}}} in~\cite{johnston2016qetlab}.

\section{D. Testing CHA and BCHA}
\label{sec:convexHull}

To approximate the set of separable states $\mathcal{S}$ with a convex hull $\mathcal{C}$,
we generate a bunch of extreme points of $\mathcal{S}$, i.e., random separable pure states in $\H_A \otimes \H_B$ in a straightforward way. The procedure for each time of sampling is demonstrated as follows:
\begin{enumerate}
\item Sample a state vector $\ket{\psi_A} \in \H_A \cong \complex^{d_A}$ from uniform distribution on the unit hypersphere in $\complex^{d_A}$, according to Haar measure ~\cite{johnston2016qetlab}.
\item Sample another state vector $\ket{\psi_B} \in \H_B \cong \complex^{d_B}$ from uniform distribution on the unit hypersphere in $\complex^{d_B}$, according to Haar measure.
\item Return $\ket{\psi_A}\ket{\psi_B}$.
\end{enumerate}

We execute the above procedure for $M$ times to gain $M$ extreme points $\textbf{c}_1, \ldots, \textbf{c}_M$.
Let $$\mathcal{C}_m := \text{conv}\left(\left\{ \textbf{0}, \ldots, \textbf{c}_m \right\}\right)$$
for $m = 1, \ldots, M$.
It is easy to see that $\mathcal{C}_m \subseteq \mathcal{C}_{m+1}$ for $m = 1, \ldots, M-1$.
Recall that we can decide whether a point $\textbf{p}$ is in $\mathcal{C}_m$ by solving the following linear programming
\begin{align}
\max \quad  & \alpha \nonumber \\
\text{s.t.} \quad & \alpha \textbf{p} = \sum_{i=0}^m\lambda_i \textbf{c}_i \nonumber,\\
\quad & \lambda_i \ge 0,\quad \sum\nolimits_{i} \lambda_i = 1.  \label{optimize}
\end{align}
If $\alpha \ge 1$, $\textbf{p}$ is in $\mathcal{C}_m$ and thus separable; otherwise, it is possibly an entangled state.
The solver for the linear programming \ref{optimize} is implemented in \href{http://qmlab.org/downloads/}{\textbf{\texttt{CompAlpha.m}}} on~\cite{qmlab}.

For the two-qubit case, we sample $M = 10^4$ extreme points, which is saved in the file \href{http://qmlab.org/downloads/}{\textbf{\texttt{2x2extreme.mat}}} on~\cite{qmlab}.
We split the data in \textbf{\texttt{2x2rdm.mat}} into two, one for training BCHA and the other for testing both CHA and BCHA.
To compare the performance of CHA and BCHA, we test the error rate of the CHA $\mathcal{C}_m$
and BCHA based on $\mathcal{C}_m$ on the test dataset.
The result is shown in table~\ref{2qubit}.
\begin{table}[hbpt]
\makeatletter
\renewcommand{\thetable}{S\@arabic\c@table}
\makeatother
\centering
{\footnotesize\renewcommand{\arraystretch}{1.5}\setlength{\tabcolsep}{5pt}
\begin{tabular}{cccccc}
\hline
$m$  		& 	1000 	& 2000 		& 3000 		& 4000 		& 5000 \\
\hline
error of CHA (\%)  	&   8.55   	& 6.01 		& 4.85  	& 4.05 		& 3.60 \\
\hline
error of BCHA (\%) 	&   3.03 	& 1.97 		& 1.47  	& 1.17 		& 1.15 \\
\hline
$m$			& 6000 		& 7000 		& 8000 		& 9000 		& 10000 	\\
\hline
error of CHA (\%)	& 3.25 		& 2.95		& 2.76		& 2.64		& 2.55	 	\\
\hline
error of BCHA (\%)	& 1.01    	& 0.75		& 0.79		& 0.71		& 0.65		\\
\hline
\end{tabular}
}
\caption{The error rate of CHA $\mathcal{C}_m$ for two-qubit separable set and BCHA based on $\mathcal{C}_{m}$ for some critical $m$.}
\label{2qubit}
\end{table}
We also apply different supervised learning algorithms with the same training and test dataset, without combining CHA. 
The result is shown in Table \ref{naive}.
\begin{table}[hbpt]
\makeatletter
\renewcommand{\thetable}{S\@arabic\c@table}
\makeatother
{\footnotesize\renewcommand{\arraystretch}{1.5}\setlength{\tabcolsep}{5pt}
\begin{tabular}{cccccc}
\hline
Method & Bagging & Boosting & SVM(rbf) & Decision Tree\\\hline
Error (\%) & 12.03 & 14.8 & 8.4 & 23.3\\
\hline
\end{tabular}
}
\caption{Error rate of classifiers trained by different algorithms. The error rate is difficult to be reduced due to the lack of prior knowledge.}
\label{naive}
\end{table}

For the two-qutrit case, we sample $M = 10^5$ extreme points, which is saved in the file \href{http://qmlab.org/downloads/}{\textbf{\texttt{3x3extreme.mat}}} on~\cite{qmlab}.
It can be verified that $66.24\%$ of the PPT random states in \textbf{\texttt{3x3rdm.mat}} are in the convex hull $\mathcal{C}_{10^5}$, which implies that
at least $66.24\%$ of the PPT random states are separable.
We used $\mathcal{C}_{10^5}$ as the criterion for separability, i.e., regard $\mathcal{C}_{10^5}$ as the true separable set.
Similar to the two-qubit case, we also tested the accuracy of CHA $\mathcal{C}_m$ as well as BCHA based on $\mathcal{C}_m$. The result is shown in table~\ref{2qutrit}.
\begin{table}[h]
\makeatletter
\renewcommand{\thetable}{S\@arabic\c@table}
\makeatother
\centering
{\footnotesize\renewcommand{\arraystretch}{1.5}\setlength{\tabcolsep}{5pt}
\begin{tabular}{cccccc}
\hline
$m$  		& 	10000 	& 20000 		& 30000 		& 40000 		& 50000 \\
\hline
error of CHA (\%)  	&  33.40   	& 22.69 		& 16.64  	&  12.60		& 9.63 \\
\hline
error of BCHA (\%) 	&   12.23 	& 9.54 		& 7.52  	& 6.07 		& 5.03 \\
\hline
$m$			& 60000 		& 70000 		& 80000 		& 90000 		&  	100000\\
\hline
error of CHA (\%)	& 6.86 		& 4.64		& 2.95		& 1.39		& 	0 	\\
\hline
error of BCHA (\%)	& 3.75    	& 2.73		& 1.81		& 1.02		& 0		\\
\hline
\end{tabular}
}
\caption{The error rate of CHA $\mathcal{C}_{m}$ for two-qutrit separable set and BCHA based on $\mathcal{C}_{m}$ for some critical $m$. Since we used $\mathcal{C}_{10^5}$ as the criterion,
the error rate of $\mathcal{C}_{10^5}$ is $0$.}
\label{2qutrit}
\end{table}

\section{E. Iterative Algorithm for Computing the Critical Point}
Recall that for an entangled state $\rho$, there exists a critical point $\alpha_{\rho}$ such that $\alpha \rho + (1 - \alpha) \mathbb{I} / (d_Ad_B)\ \; (0 \le \alpha \le 1)$ is separable
when $\alpha \le \alpha_{\rho}$ and entangled when $\alpha > \alpha_{\rho}$.
Based on CHA, we developed an iterative algorithm for approximating $\alpha_{\rho}$ in a more efficient way, which is shown as follows:
\begin{enumerate}
\item Randomly sample $1000$ extreme points and form a convex hull $\mathcal{C}$.
Let $\textbf{p}$ be the feature vector of $\rho$. Set $\epsilon = 1$, $\gamma = 0.95$.
\item
Update $\alpha_{\rho} \leftarrow \alpha(\mathcal{C}, \textbf{p})$.
\item Suppose now $\mathcal{C} := \text{conv}\left(\left\{ \textbf{c}_1, \ldots, \textbf{c}_m \right\} \right)$, and $\alpha_{\rho}\textbf{p} = \sum_{i} \lambda_i \textbf{c}_i$.
Let $\textbf{c}_{i_1}, \ldots, \textbf{c}_{i_D}$ be the extreme points such that $\lambda_{i_k} > 0$. Set $\mathcal{C} \leftarrow \text{conv}\left(\left\{ \textbf{c}_{i_1}, \ldots, \textbf{c}_{i_D}\right\}\right)$.
\item For each $k = 1, \ldots, D$, suppose $\textbf{c}_{i_k}$ is the feature vector of $\ket{a_k}\ket{b_k}$.
We randomly generate two Hermitian operators $H_1 \in \text{End}(\mathcal{H}_A)$, $H_2 \in \text{End}(\mathcal{H}_B)$ such that $\norm{H_1}_2 = 1$ and $\norm{H_2}_2 = 1$.
Let $\xi$ be a random number in $[0, \epsilon]$. Set $\ket{a_k'}\ket{b_k'} = \left(e^{i\xi H_1} \otimes e^{i\xi H_2}\right) \ket{a_k}\ket{b_k}$.
Set $\mathcal{C} \leftarrow \text{conv}\left(\left\{ \mathcal{C}, \textbf{c}_k'\right\}\right)$, where $\textbf{c}_k'$ as the feature vector of $\ket{a_k'}\ket{b_k'}$.
\item Set $\epsilon \leftarrow \gamma\epsilon$ and back to step 2.
\end{enumerate}

What step 4 does is sampling in the neighborhood of $\textbf{c}_{i_k}$. In practice, we repeat step 4 for $10$ times to get a bunch of neighbors.
The detailed implementation is in the code \textbf{\texttt{CriticalPoint.m}} on ~\cite{qmlab}.

\end{document}